\documentclass[draft]{agujournal2019}
\usepackage{url} 
\usepackage{lineno}
\usepackage[inline]{trackchanges} 
\usepackage{soul}
\usepackage{siunitx}
\usepackage{gensymb}
\usepackage{float}
\usepackage{soul}

\draftfalse

\journalname{QJRMS}

\begin{document}
\title{An observationally constrained probabilistic trigger for organized deep convection in an NWP ensemble}

\authors{M. R. Muetzelfeldt\affil{1},
R. S. Plant\affil{1},
H. M. Christensen\affil{2},
Z. Zhang\affil{2},\\
T. Woollings\affil{2},
A. Stirling\affil{3},
W. Tennant\affil{3}, and
M. Moncrieff\affil{4}
}

\affiliation{1}{Department of Meteorology, University of Reading, Reading, UK}
\affiliation{2}{Atmospheric, Oceanic and Planetary Physics, University of Oxford, Oxford, UK}
\affiliation{3}{Met Office, Exeter, UK}
\affiliation{4}{Climate and Global Dynamics, National Center for Atmospheric Research, USA}

\correspondingauthor{Mark Muetzelfeldt}{mark.muetzelfeldt@reading.ac.uk}

\begin{keypoints}
\item A novel stochastic scheme is used to represent MCS activity in atmospheric models
\item The trigger condition is constrained by environmental conditions associated with organized deep convection
\item The scheme is shown to improve the spread-error relationship in tests with an NWP ensemble
\end{keypoints}

\begin{abstract}
A novel stochastic parametrization scheme representing organized convection is described. The effects of mesoscale convective systems (MCSs) are represented in an observationally constrained manner, by probabilistically triggering an MCS scheme in regions of enhanced environmental total column water vapour. In combination with the probabilistic trigger, patterns with given spatiotemporal scales determine where and when the scheme is active. Our scheme builds on the multiscale coherent structure parametrization (MCSP), which represents the top-heavy heating structure associated with MCSs.

The original and new MCSP schemes are tested in a numerical weather prediction (NWP) ensemble. Both MCSP schemes improve the spatiotemporal scales of tropical precipitation compared to a control. When the spread-error relationship of tropical precipitation is analysed, the new scheme successfully boosts spread compared to original MCSP, improving the underdispersion of the ensemble seen with the original MCSP.
\end{abstract}

\section{Introduction}

Atmospheric convection can organize at many different scales. Mesoscale convective systems (MCSs) represent an important category of organized convection, producing high fractions of total rainfall over large swathes of the tropics and extratropics \cite{nesbitt2006storm,feng2021global}. They can be both mesoscale collections of many deep convective cumulonimbus with accompanying anvil cloud, as well as constituents of larger-scale organized convection such as convectively coupled equatorial waves (CCEWs), the Madden-Julian Oscillation (MJO), and the InterTropical Convergence Zone (ITCZ). Their anvil clouds distinguish them from simply being aggregations of cumulonimbus clouds, by altering their heating profile to be more top-heavy \cite{schumacher2004tropical,houze2004mesoscale}. Here, we investigate the effect of adding a top-heavy heating profile to a numerical model, doing so in an observationally constrained way based on the probability of MCS occurrence for the first time.

Relatively little direct attention has been paid to how to represent organized convection in coarser models, instead relying on the interaction between the model's convection parametrization and microphysics schemes to create realistic approximations of MCSs. However, this may not be sufficient. The mesoscale circulations that occur within MCSs cannot be represented by models with a grid length of greater than 20 km, and biases have been seen at grid lengths as fine as 10 km \cite{moncrieff2006representing}. Here, we extend a scheme which is designed to represent the missing top-heavy heating caused by MCSs, the multiscale coherent structure parametrization \cite <MCSP;> {moncrieff2017simulation}. In models with grid lengths of order 100-km, this has been shown to improve precipitation in the ITCZ and the coupling to CCEWs, as well as improving the MJO \cite{moncrieff2017simulation,chen2021effects,zhang2025advancing}. \citeA{zhang2025advancing} also investigated how MCSs are represented in a 75-km (grid spacing over tropics) model using MCSP, and we build on this work.

To know when to apply MCSP, it is necessary to have a model of when upscale organization of convection will occur. In previous studies, this has come solely from the convection parametrization scheme and the dynamical wind conditions, relying on the presence of convection along with a minimum threshold on wind shear to trigger the scheme \cite{moncrieff2017simulation,chen2021effects,zhang2025advancing}. We augment these trigger conditions based on recent work that shows that the probability of MCS-type convection occurring relative to all deep convection scales approximately linearly with total column water vapour \cite <TCWV;> {muetz2025environmental}. The aim is to ensure that MCSP is active only in those places where the environment is favourable for MCS formation; a more targeted observationally constrained approach to triggering the scheme.

The probabilistic nature of our targeted trigger lends itself to an approach linked to existing stochastic parametrization schemes. Schemes such as stochastic perturbed parameter tendencies \cite <SPPT;> {buizza1999stochastic,berner2017stochastic} use a generated pattern with given decorrelation length and time scales to modulate their activity. Our approach is to use the generated pattern as a probability field, determining the likelihood of MCS occurrence by testing against a probability calculated from the environmental conditions. This represents a new class of phenomenon-based stochastic parametrization, distinct from model-uncertainty \cite <e.g.,> {buizza1999stochastic} and sampling-based \cite <e.g.,> {plant2008stochastic} stochastic parametrization. The pattern field has a suitable length and time scales to represent MCSs. Unlike SPPT, which can only scale the structure provided by existing parametrizations, our new scheme can represent new physics, the organization of convection, with a changed structure to the column heating through the introduction of a top-heavy heating profile.

To explore the effect of our scheme, we use an NWP ensemble. Our hypothesis is that adding a stochastic trigger will increase the spread of our simulations compared to MCSP only, similar to existing stochastic parametrization schemes \cite{palmer2009stochastic}. However, our scheme differs from existing ones by first having a strong link to an atmospheric phenomenon, MCSs, and second providing some justification for the length and time scales associated with the generated pattern. We initialize forecasts on the first of each month spanning one year, ensuring that we sample different seasons and atmospheric conditions to test the robustness of our results.

The aim of this work is to test our new scheme in an ensemble setting, testing first the spatial and temporal scales of precipitation, before examining the spread-error relationships of the ensemble. In Section \ref{sec_methods}, we set out our data sources and describe the existing MCSP scheme, before detailing our modifications to it and our analysis methods. We present our results in Section \ref{sec_results}, showing how our scheme affects the spatiotemporal scales of precipitation, before analysing ensemble spread-error characteristics. We end with a conclusion and some closing remarks in Section \ref{sec_conclusions}.

\section{Data and Methods}
\label{sec_methods}

\subsection{GPM IMERG Precipitation}

For comparison against observed precipitation, we use the Global Precipitation Measurement (GPM) Integrated Multi-satellitE Retrivals for GPM (IMERG) Final product (version 6), which provides global estimates of precipitation at half-hourly temporal frequency over a spatial discretization of 0.1\degree~\cite{huffman2020IMERG}. The Final product includes a monthly mean correction to gauge data where available. We use the data at hourly temporal frequency to match our simulation output, and we coarsen the data to the resolution of our model using conservative interpolation.

%

\subsection{Ensemble Simulations}

We use N216 resolution Unified Model (UM) simulations (approximately \SI{75}{km} grid length at the equator), as described in \citeA{zhang2025advancing}, with 10 ensemble members. Ensemble member 0 is a deterministic control, so we disregard results from this member. We run with three model configurations: \textit{Control}, \textit{PRIME-MCSP}, and \textit{Stoch-PRIME-MCSP}. \textit{Control} uses the standard operational UM settings \cite{bush2023second} with the addition of the CoMorph-A convection parametrization scheme \cite{lock2024performance}, which all simulations use due to its improved representation of precipitation temporal variability. \textit{PRIME-MCSP} is original version of MSCP exactly as described in \citeA{zhang2025advancing}, and \textit{Stoch-PRIME-MCSP} is the stochastic version of MCSP described below. For each configuration, 12 simulations initiated on the 1\textsuperscript{st} of each month of 2020 are run for 10 days, with the initial conditions provided by MOGREPS-G \cite{inverarity2023met}, ensuring that we have simulated days that span all seasons. We use both stochastic physics schemes in \textit{Control}: stochastic kinetic energy backscatter \cite <SKEB:> {tennant2011using} and stochastic perturbed tendencies \cite <SPT:> {sanchez2016improved} to maintain a similar configuration to the operational MOGREPS-G ensemble. These are also enabled in the MCSP simulations to mimimize the configuration changes between these and \textit{Control}.

\subsection{Multiscale Coherent Structure Parametrization}

MCSP was developed to represent the heating and momentum effects of organized deep convection \cite{moncrieff2017simulation}. The key argument behind its development is that similar coherent structures resembling the archetypal squall line occur over a wide range of scales. MCSP parametrizes the effects of such systems by adding a top-heavy heating tendency and a momentum tendency to those produced by a standard convection parametrization scheme. Top-heavy heating profiles are associated with organized deep convection and MCSs \cite <e.g.>{houze2004mesoscale}. In \citeA{zhang2025advancing} we developed a simple improved trigger for the MCSP scheme to better represent MCSs: the top-heavy heating is triggered by the presence of deep convection, which must have a depth of over 300 hPa and a top above the freezing level, as well as a shear of over \SI{3}{m.s^{-1}} between \SI{5}{km} and \SI{500}{m}. We call this the \textit{PRIME-MCSP} scheme.

\subsubsection{MCSP Observationally Constrained Probabilistic Trigger}

We extend \textit{PRIME-MCSP} in two ways. First, we add in extra dependence on the environmental conditions in the model. To do this, we make use of the observed relationship between total column water vapour (TCWV) and MCS occurrence discovered by \citeA{muetz2025environmental}. We found that the probability of MCS-type convection given that deep convection is active scales close to linearly with TCWV (our Figure 8a). The knowledge of whether deep convection is active in the numerical model is provided by the convection parametrization scheme, and so this relationship gives us the probability that the parametrized convection should be MCS-like based on the grid-column value of TCWV. The relationship can thus be used as the basis for a targeted trigger for MCSP.

The relationship from \citeA{muetz2025environmental} is inherently probabilistic. To apply it, we combine the conditional probability of MCS-type convection with the pattern generated by stochastic parametrization schemes (described below). These patterns are an integral part of stochastic parametrizations, and ensure that the applied stochasticity is not filtered out by the model's dynamics \cite{kober2016physically}. They are used to represent model error, and increase ensemble spread such that the spread-error relationship holds more closely \cite{palmer2009stochastic}. The patterns have a characteristic length and time scale associated with them, and we want information about the presence or absence of an MCS to span multiple grid cells, so using these patterns is a natural fit.

The stochastic patterns used for the operational Met Office global ensemble have a decorrelation length scale set to \SI{500}{km}, and decorrelation time scale set to approximately \SI{6}{h} (\SI{20000}{s}) \cite{inverarity2023met}. These have been tuned to give the best results for the stochastic schemes used by the Unified Model, the SPT and SKEB schemes \cite{sanchez2016improved}. Since these scales are broadly similar to the length and time scales associated with MCSs \cite <e.g.,>{feng2021global}, we adopt them here as being representative of the scales of organized deep convection. However, we apply a transform to the pattern -- a translation by 180\degree~about the pole\footnote{This is not a rotation: each point is effectively translated by half a circumference along its meridian, such that the transform does not retain any correlation with the original pattern.} -- so that the pattern we use is uncorrelated with those used by SPPT and SKEB. We call the new scheme developed here \textit{Stoch-PRIME-MCSP} due to its use of stochastic patterns. Halving or doubling the decorrelation length scale did not have a substantial impact on the spatial scale of simulated precipitation features (Buppanimitr, 2025, pers. comm.).

\subsection{Analysing Scales of Precipitation}
\label{sec_ASoP}

To assess the impact of our modifications of convection parametrization, we use analysis based on Analysing Scales of Precipitation \cite <ASoP;>[cf their Figures~1 and 2] {klingaman2017ASoP}. Specifically, we partition the rainfall into intensity bins in order to explore an existing bias in the original MCSP scheme, whereby too much precipitation is seen in the western Pacific \cite{zhang2025advancing}. The analysis is performed by calculating the fractional contributions of different precipitation bins, low, moderate, heavy and extreme rainfall (see Figure \ref{fig2_precip_contrib} for the bin ranges).

We also compute spatiotemporal correlations of rainfall to explore how the MCSP schemes match the scales of precipitation associated with MCSs. This analysis involves first calculating a 3-hour mean of the precipitation data, then calculating the correlation between a reference grid cell and those surrounding it in the $x$- and $y$-dimensions. Correlations are also performed along the time dimension. The information from the spatial dimensions is aggregated from those cells which are a given distance from the reference grid cell (Figure \ref{fig3_spat_temp_corr}).

\subsection{Gaussian Smoothing of Precipitation}
\label{sec_gaussian_smooth}

Precipitation is a noisy field with high variance. To perform spread-error analysis on such a field, it can be useful to apply spatial smoothing . This enables one to avoid the ``double counting'' problem for small displacements of localized rainfall features. For regional modelling, a common technique is to use fractions skill score \cite <FSS;>{roberts2008assessing}, which can be adapted to produce spread-error relationships \cite{dey2014spatial}. However, such an analysis is tied to the underlying grid, and cannot easily take into account cells with aspect ratios that differ from unity, as in our simulations. Therefore, we prefer instead to use a Gaussian smoothing of the precipitation to facilitate spread-error analysis. Gaussian smoothing is applied over the entire tropics (30\degree S--30\degree N) using various length scales to obtain a series of spread-error relationships as a function of the width of the smoothing kernel. The aspect ratio is a function of latitude, as the zonal distance changes whereas the meridional does not for a latitude-longitude grid; we use the mean aspect ratio over the tropics to scale the smoothing kernel.

We use root mean squared error (RMSE) as our error metric. Adapting \citeA{dey2014spatial} who used similar definitions for FSS, we compute a dispersion RMSE (dRMSE) to characterize the spread of an ensemble, as measured by the mean of the pairwise member-to-member RMSE at a given spatial smoothing. We similarly compute an ensemble RMSE (eRMSE) to characterize the error, which is measured by the mean of the member-to-observation RMSE at a given spatial smoothing. 

\section{Results}
\label{sec_results}

\subsection{Spatiotemporal Scales of Precipitation}
\label{sec_spatiotemporal_scales_of_precip}

\begin{figure}[H]
\centering
\includegraphics[width=\textwidth]{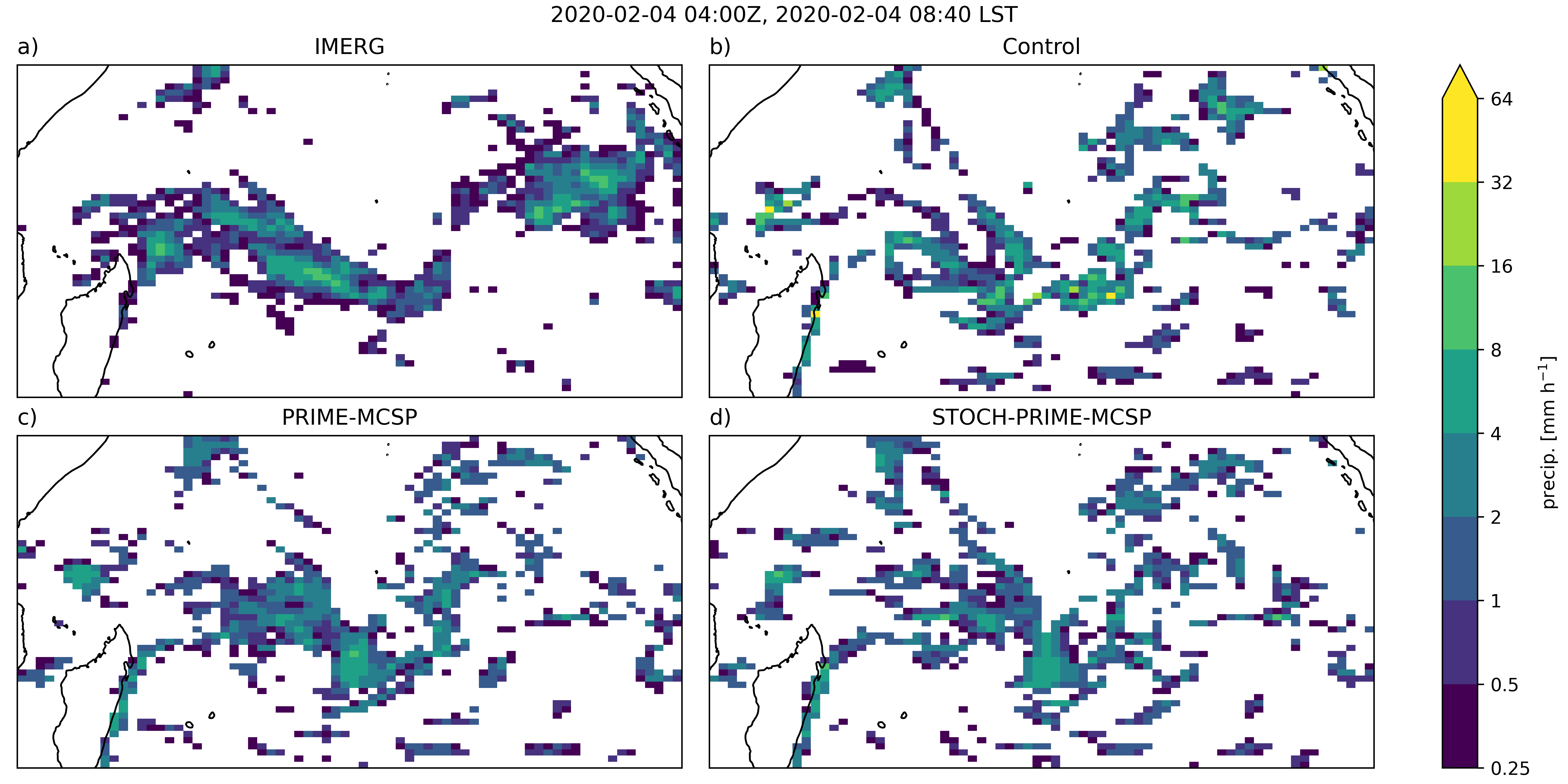}
\caption{Instantaneous precipitation rates over the Indian Ocean in (a) IMERG coarsened to the resolution of the model grid and (b) \textit{Control}, (c) \textit{PRIME-MCSP}, and (d) \textit{Stoch-PRIME-MCSP} simulations at 08:40 local solar time (LST; from centre of domain) on 4 February 2020.}
\label{fig1_4xprecip_over_IO}
\end{figure}

To illustrate the nature of convection in the three simulations, we show a snapshot of precipitation over the Indian Ocean in Figure \ref{fig1_4xprecip_over_IO}. Figure \ref{fig1_4xprecip_over_IO}b shows that the rainfall in \textit{Control} is too broken up, and not aggregated enough, compared to IMERG. Where precipitation occurs, it tends to be too strong. \textit{PRIME-MCSP} does better at capturing the organized convection near the centre left of the domain, but is too broken on the eastern side (Figure \ref{fig1_4xprecip_over_IO}c). \textit{Stoch-PRIME-MCSP} is similar to \textit{PRIME-MCSP}, with more aggregation than \textit{Control} but with perhaps too many localized centres of scattered convection (Figure \ref{fig1_4xprecip_over_IO}d). The snapshot is one day into the simulation, so it is not expected that the simulations will exactly reproduce the organized features, but these examples are indicative of the general characteristics of each model. These results suggest that there is not enough spatial organization of convection in \textit{Control}.

\begin{figure}[H]
\centering
\includegraphics[width=\textwidth]{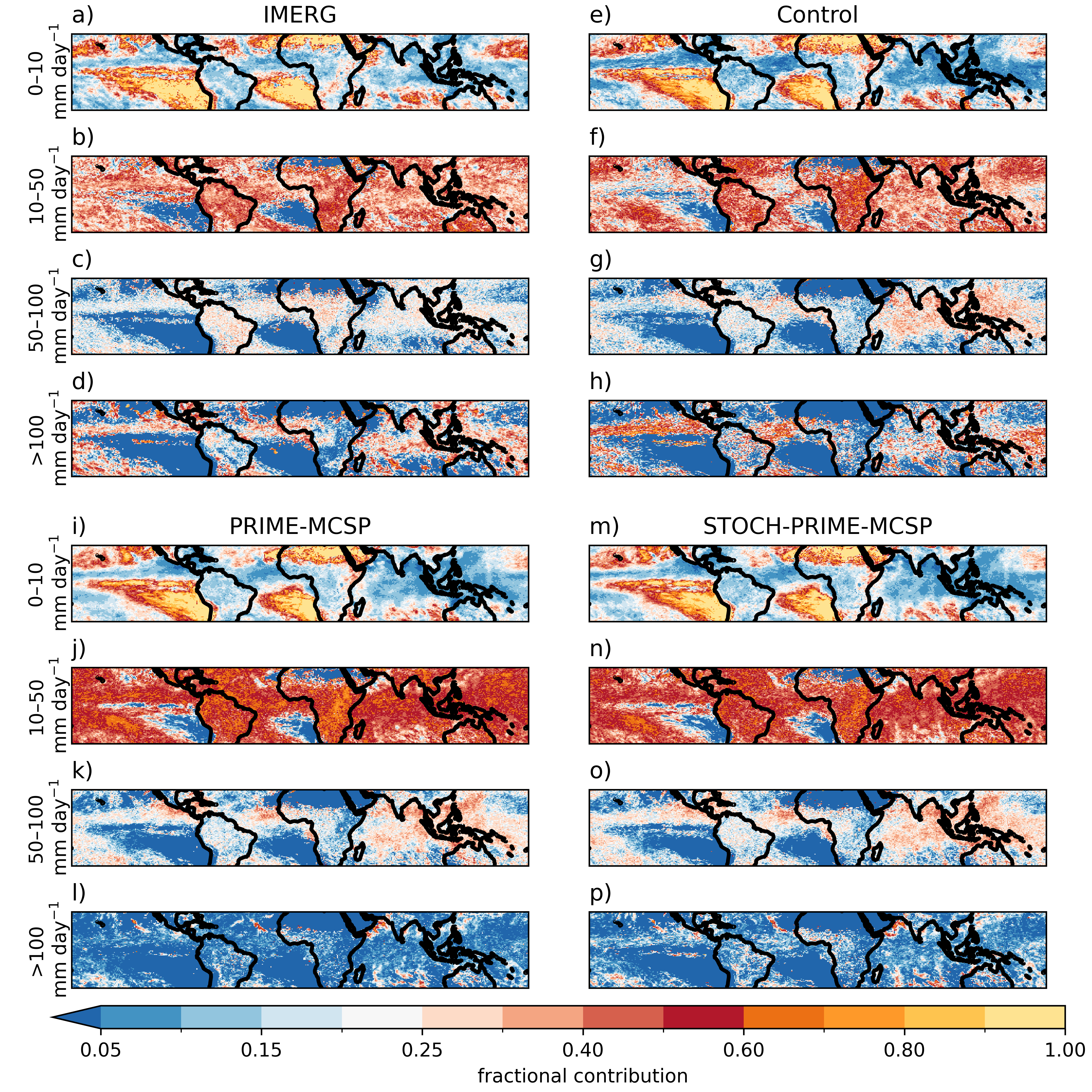}
\caption{Fractional contribution of binned rate of precipitation to the total over the tropics for IMERG (four top-left panels) and \textit{Control}, \textit{PRIME-MCSP} and \textit{Stoch-PRIME-MCSP} (one per quadrant). The bin ranges are above each panel, and are referred to in the text as low, moderate, heavy, and extreme. The sum for each grid cell in a four-panel quadrant is unity.}
\label{fig2_precip_contrib}
\end{figure}

\begin{figure}[H]
\centering
\includegraphics[width=0.8\textwidth]{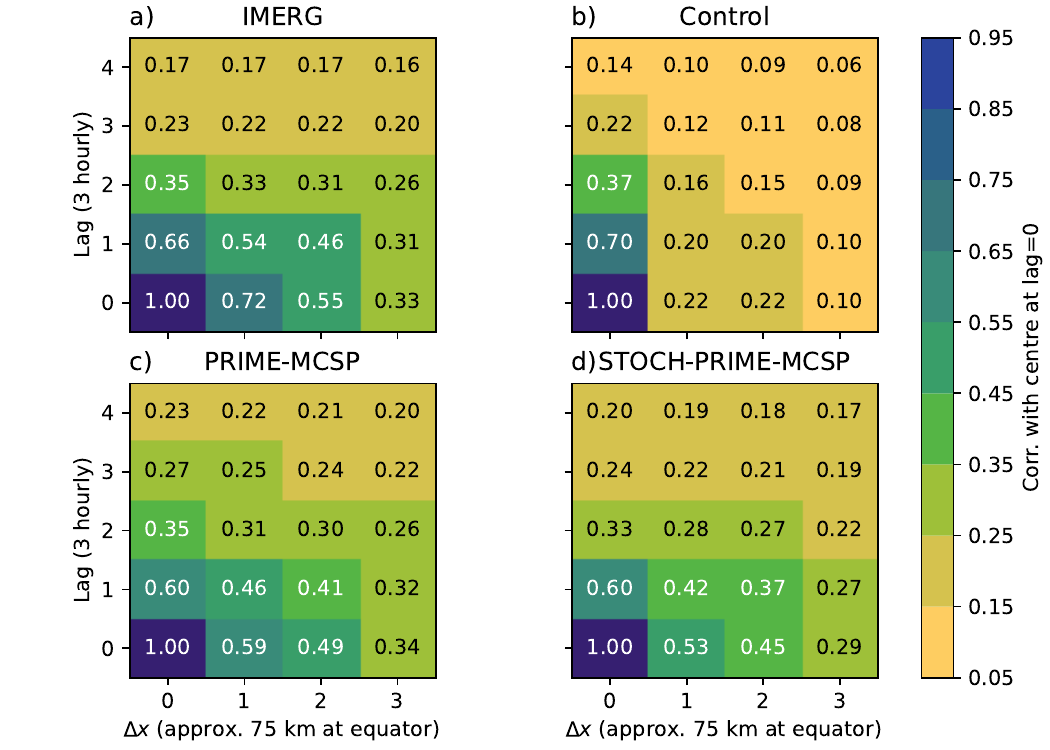}
\caption{Spatiotemporal correlation of precipitation over the tropics for (a) IMERG and (b) \textit{Control}, (c) \textit{PRIME-MCSP}, and (d) \textit{Stoch-PRIME-MCSP} simulations. The spatial correlation of each grid cell at a given distance from the reference grid cell is shown in the $x$ direction, and the temporal correlation of each grid cell with the reference grid cell is shown in the $y$ direction. A 3-hour time average is used, as in \citeA{klingaman2017ASoP}.}
\label{fig3_spat_temp_corr}
\end{figure}

We assess the degree of spatiotemporal organization using ASoP analysis (Section \ref{sec_ASoP}). The results are from a single, non-deterministic ensemble member, encompassing the full time span of all the simulations: 12 start dates for 10 days each. The results of the other ensemble members were checked, and all show the same qualitative behaviour with little variation between them -- the same regional patterns for Figure \ref{fig2_precip_contrib} and typically within one one-hundredth of the absolute difference in correlations for Figure \ref{fig3_spat_temp_corr} (not shown).

We investigate how well the different simulations match the precipitation from different rain rates in Figure \ref{fig2_precip_contrib}: we show the fractional contribution of the total for each simulation that precipitation rates (expressed as \si{mm.day^{-1}}) occupy of each of the four bins. For IMERG, Figure \ref{fig2_precip_contrib}a,b,c,d can be compared to \citeA{klingaman2017ASoP} (their Figure 1 is only for the equatorial warm pool, see our Figure S1 for a direct comparison). These show similar partitioning between the 4 bins, with the moderate bin providing around 0.4--0.6 of the total precipitation over the warm pool. Extreme precipitation ($>$\SI{100}{mm.day^{-1}}) provides an important contribution over the warm pool, Indian Ocean, and ITCZ.

\textit{Control} (Figure \ref{fig2_precip_contrib}e,f,g,h) has too little contribution from the low bin over the western Pacific subtropical high, and too much from the moderate bin in this region. The heavy bin provides too much precipitation across the ITCZ, Indian Ocean, and warm pool.

Both \textit{PRIME-MCSP} and \textit{Stoch-PRIME-MCSP} show the moderate bin providing too much rainfall (0.5--0.7) over most areas except the subtropical highs and the Sahara. Both contribute too little from the low and extreme bins. The heavy bin also contributes too much, and more so than \textit{Control}, particularly over the western Pacific. This sheds light on the overestimation of precipitation in this region with \textit{PRIME-MCSP} \cite{zhang2025advancing}, pointing to too much heavy precipitation over the western Pacific in the simulations, although they were using a coarser climate model than here. The heavy rain \textit{rates} indicate that this is likely to be driven by organized systems, with both the convective and stratiform components contributing. Figure \ref{fig2_precip_contrib}g shows too much heavy precipitation over the Indian Ocean in \textit{Control}, a region where there is a known bias in the UM \cite{bush2015effect} and which \textit{PRIME-MCSP} improved \cite{zhang2025advancing}. The mechanism here seems to be the same as for the western Pacific but in reverse; MCSP schemes improve the bias here by reducing the heavy, presumably organized convection in this region.

Figure \ref{fig3_spat_temp_corr} shows the spatiotemporal correlation of precipitation at the grid scale \cite{klingaman2017ASoP}. Figures~\ref{fig3_spat_temp_corr}a,b show that \textit{Control} is poorly correlated in space with a reference grid cell, but well correlated in time. The poor spatial correlation is consistent with the reduced organized convection seen in Figure \ref{fig1_4xprecip_over_IO}. \textit{PRIME-MCSP} and \textit{Stoch-PRIME-MCSP} both improve the spatiotemporal correlations  to the reference cell. \textit{PRIME-MCSP} produces the most realistic spatial correlation, although it is not strong enough at the shortest spatial scale. There is little difference between \textit{PRIME-MCSP} and \textit{Stoch-PRIME-MCSP} for the temporal correlation, although the temporal correlation of \textit{PRIME-MCSP} is marginally too strong at longer lag times of nine hours (Lag 3) and longer.

These results indicate that the MCSP schemes improve the organization of convection in these simulations compared to \textit{Control}, with improved spatiotemporal correlations between grid cells (Figure \ref{fig3_spat_temp_corr}). A plausible mechanism for this could be that the top-heavy heating structure triggers a larger-scale adjustment, for example by a gravity wave response, that organized the convection and its precipitation on larger spatial scales. However, the MCSP schemes do degrade the overall distributions of precipitation (Figure \ref{fig2_precip_contrib}) compared to \textit{Control}. The top-heavy heating profile will stabilize the troposphere, which might lead to too little precipitation in the extreme bin as the convection decreases. This would lead to a greater fraction of precipitation coming from the moderate and heavy bins.

\subsection{Comparing Spread and Error}

\begin{figure}[H]
\centering
\includegraphics[width=\textwidth]{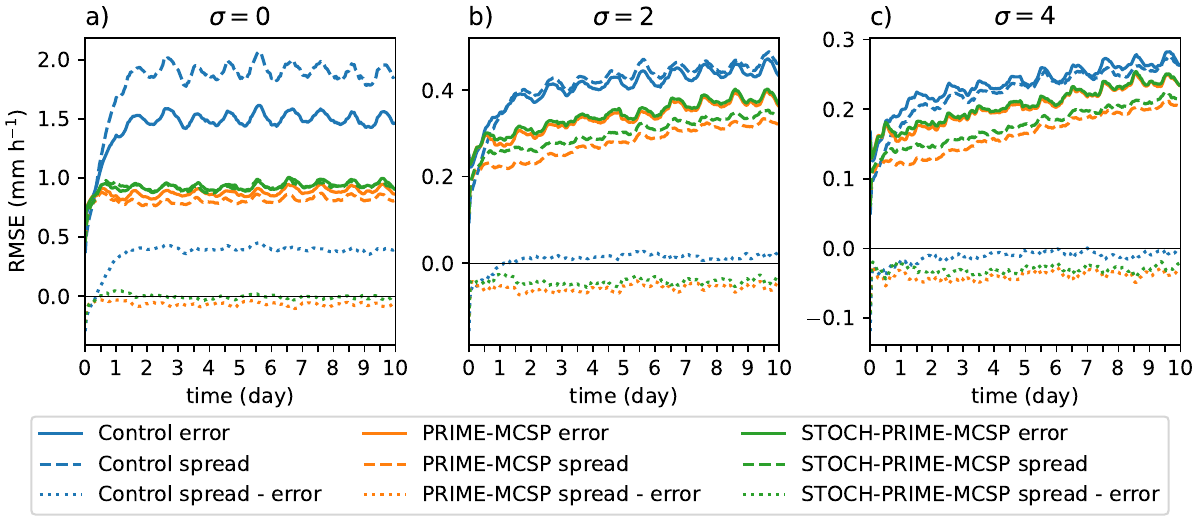}
\caption{Precipitation spread and error over the tropics for the three configurations, at three scales of spatial smoothing ($\sigma$). Spread is given by dRMSE (dashed lines), and error by eRMSE (solid lines; Section \ref{sec_gaussian_smooth}). (a) $\sigma = 0$ is no spatial smoothing (i.e., at the grid scale), (b) $\sigma = 2$ applies Gaussian smoothing with a kernel that has an $e$-folding distance of 2 times the grid spacing in the zonal direction, and similarly (c) for $\sigma = 4$. The dotted lines gives spread minus error; if it is above zero this means the simulation is overdispersed and \textit{vice versa}. Data from all 12 start dates are averaged in each case. }
\label{fig4_spr_err_precip_scales}
\end{figure}

Does the improvement of the structure of organized precipitation lead to an improvement in ensemble skill? We address this question with Figure \ref{fig4_spr_err_precip_scales}, which compares the temporal growth of ensemble spread and error in the ensemble mean for precipitation forecasts. The three panels show different levels of smoothing of the precipitation fields. Figure \ref{fig4_spr_err_precip_scales}a shows that \textit{Control} is overdispersed for unsmoothed precipitation forecasts for all times beyond 12 h. \textit{PRIME-MCSP} and \textit{Stoch-PRIME-MCSP} both have much lower errors than \textit{Control}, indicating much improved skill. The spread is higher for \textit{Stoch-PRIME-MCSP}, as hypothesized due to adding in stochasticity, and for no spatial smoothing almost exactly matches the error. \textit{PRIME-MCSP} is slightly underdispersed. 
A diurnal cycle is evident for all configurations, with the likely cause being the distribution of land masses in the tropics. The peak occurs at approximately 1300Z, indicating that the strong diurnal cycle over Africa could be responsible, as 1300Z is close to when solar forcing is at a maximum and when the simulated diurnal cycle of convection over land typically peaks \cite <e.g.,>{muetz2021evaluation}.

With no spatial smoothing, the error for each simulation saturates at around day 2 (notwithstanding the diurnal cycle), indicating that no useful forecast information is being generated at the grid scale by the models after this time, as the simulations have developed into a climatology, or are potentially capturing the slow variation of tropical models like the MJO and El Niño-Southern Oscillation (ENSO), of precipitation. Some sort of spatial smoothing (e.g., FSS, see Section \ref{sec_gaussian_smooth}) is advisable for comparing models and observation at the grid-scale, as NWP models do not have much skill for precipitation at this scale, and it avoids the double counting problem. The MSCP simulations are plausibly smoothing the precipitation field because of their use of spatiotemporal correlated patterns or their better representation of larger organized features, increasing their skill at this level.

When spatial smoothing is applied (Figures \ref{fig4_spr_err_precip_scales}b,c),  \textit{Control} has the best spread-error relationship. However, the error for both MCSP simulations remains lower than that for \textit{Control}, again showing an improvement in model skill. Spread for \textit{Stoch-PRIME-MCSP} is consistently higher than that for \textit{PRIME-MCSP}, demonstrating the value of including stochasticity in the trigger function and matching our initial hypothesis. The spread and error increase (with diurnal cycle superimposed) over the 10-day duration of these experiments, and have not saturated by the end. This indicates that there is substantial noise in the unsmoothed precipitation which potentially saturates the spread and error earlier for them than for the smoothed precipitation.

That adding in MCSP (\textit{PRIME-MCSP}) decreases the spread is interesting. This implies that adding in a top-heavy heating where there is convection makes the upscale growth of \textit{differences} between ensemble members smaller. We speculate that this could be due to a more consistent upscale growth of convection between the different ensemble members, i.e. an improvement in the coupling between convection and organization. This may mean that if \textit{PRIME-MCSP} were to be run operationally, it would be necessary to increase the strength of the stochastic parametrizations to counteract this reduction in spread.

Overall, adding in MCSP is positive for model skill, and \textit{Stoch-PRIME-MCSP} further improves the spread-error relationship. Given the results from Section \ref{sec_spatiotemporal_scales_of_precip} indicating that both MCSP schemes improve spatiotemporal precipitation scales, we interpret this as indicating that the improved spatiotemporal structures lead to an improvement in model skill. This implies that the organized convection that MCSP represents is improving not just the structure of the precipitation field (Section \ref{sec_spatiotemporal_scales_of_precip}), but the upscale effect that the organized convection produces improves the location of larger-scale precipitation features. At a coarser resolution, \textit{PRIME-MCSP} improves tropical phenomena such as Kelvin waves and the MJO \cite{zhang2025advancing}, and it is possible that these improvements are present at the resolution used here. If so, this could also lead to an improvement in skill over longer time periods. However, it is difficult to assess this with only 10-day simulations.

\section{Conclusions}
\label{sec_conclusions}

We have developed a new class of phenomenon-based stochastic parametrization, designed to represent organized deep convection in this case. It triggers MCSP selectively where MCS activity is likely, using the background TCWV field as an environmental control. It is different from stochastic parametrizations such as SPPT because it can have impacts by triggering new behaviour, not just modifying tendencies produced by parametrization schemes but changing their vertical structure. However, similarly to SPPT, it does require the existing convection scheme to be active in order for it to trigger, and it is possible for it to have a secondary effect on subsequent triggering by feedbacks onto the larger-scale state. \textit{Stoch-PRIME-MCSP} uses the spatiotemporal correlation patterns from SPPT, which are reasonably well suited to representing MCSs. Improving the evidence base for choosing the length and time decorrelation scales for the stochastic pattern fields and performing experiments with changes to these are promising avenues for future research.

Both MCSP schemes improve the organization of convection in our simulations, producing more realistic spatial patterns of precipitation and better near-grid-scale correlations compared to IMERG. This indicates that the top-heavy heating profile produced by these schemes is beneficial for organizing convection, supporting the rationale in \citeA{moncrieff2017simulation}, presumably through feedback onto the model's dynamics. Figure \ref{fig2_precip_contrib} indicates that the overestimation of precipitation by MCSP in the western Pacific \cite{zhang2025advancing} is due to the heavy rain produced by these simulations. This should be explored further by more closely analysing this region by, for example, varying the strength of the top-heavy profile and exploring how this affects the precipitation response in the western Pacific and the precipitation distribution globally.

The MCSP configurations both improve the error of simulated precipitation compared to IMERG, perhaps as a consequence of the improvement in organized convection. However, \textit{PRIME-MCSP} is underdispersed even as the precipitation field is smoothed to allow comparisons at larger scales. \textit{Stoch-PRIME-MCSP} improves this underdispersion by boosting spread, consistent with our original hypothesis. Both MCSP schemes therefore provide value at improving forecasts out to 10 days -- extending the simulations would allow these results to be tested further.

Here, we only treat the top-heavy heating from MCSs. In previous a previous study, both the heating and momentum transports have been considered \cite{moncrieff2017simulation}, and the momentum was shown to affect the precipitation structure. In future work, MCS momentum transports should also be considered, with care taken to calculate the propagation direction and MCS orientation in order to correctly apply the momentum transport. Additionally, it may be necessary to categorize shear-perpendicular and shear-parallel MCSs, which have different characteristics for momentum transport, and are particularly important in the ITCZ.

Other phenomena may be candidates for similar schemes. Here, we have focused on MCSs when they are of a similar size to the grid spacing, i.e. we are in the MCS grey zone. A potential candidate for a follow-on scheme would be representing individual cumulonimbus clouds in a kilometre-scale model, in the deep convective grey zone.

\section*{Statement of Author Contributions}

MRM: Conceptualization, Investigation, Methodology, Software, Validation, Visualization, Writing -- original draft.
RSP: Methodology, Funding acquisition, Writing -- review and editing.
HMC: Methodology, Funding acquisition, Writing -- review and editing.
ZZ: Methodology, Writing -- review and editing.
TW: Methodology, Funding acquisition, Writing -- review and editing.
AS: Methodology, Writing -- review and editing.
WT: Methodology, Writing -- review and editing.
MM: Writing -- review and editing.

\section*{Conflict of Interest}

The authors declare no conflicts of interest.

\section*{Acknowledgements}

All simulations were run on Monsoon2, a collaborative High-Performance Computing facility funded by the Met Office and the Natural Environment Research Council \url{https://www.metoffice.gov.uk/research/approach/collaboration/jwcrp/monsoon-hpc}. All analysis was done on JASMIN, the UK's collaborative data analysis environment (\url{https://www.jasmin.ac.uk}).

\section*{Funding Information}

This work was made possible by the Mesoscale Convective Systems: PRobabilistic forecasting and upscale IMpacts in the grey zonE (MCS:PRIME) Natural Environmental Research Council (NERC) Grant NE/W005530/1, which supported MRM, RSP, HMC, ZZ, and TW. HMC was also supported by NERC grant NE/P018238/1 and by a Leverhulme Trust Research Leadership Award  `Seamless Uncertainty Quantification for Earth System prediction' (SUQCES).

\section*{Data Availability}

The Unified Model output is stored on JASMIN, and will be made available upon reasonable request.
The Unified Model source code is available to use under licence. The specific branch and configuration used for these simulations can be found at \url{https://code.metoffice.gov.uk/trac/um/browser/main/branches/dev/markmuetzelfeldt/vn13.0_org_conv} (Met Office Science Repository Service (MOSRS) account required).

The GPM IMERG Final Precipitation Half Hourly 0.1 degree x 0.1 degree V06 data were obtained from the NASA Goddard Earth Sciences Data and Information Services Center (GES DISC) at \url{https://doi.org/10.5067/GPM/IMERG/3B-HH/06}.

\section*{Code Availability}

All analysis scripts used to produce the figures in this paper are available at \url{https://github.com/markmuetz/MCSP_stoch_trigger_ensemble}, and will be archived before final submission. 

\bibliography{references}

\end{document}